\documentclass[aps,pra,
epsfigure,twocolumn,longbibliography,superscriptaddress]{revtex4-2}
\usepackage{amsfonts}
\usepackage{amssymb}
\usepackage{amsmath}
\usepackage{calc}
\usepackage{subfigure}
\usepackage{graphicx}
\usepackage{epstopdf}
\usepackage{dcolumn}
\usepackage{bm}
\usepackage{color} 
\usepackage{txfonts}
\usepackage[dvipsnames]{xcolor}
\usepackage{appendix}
\usepackage[colorlinks, citecolor=blue]{hyperref} 
\usepackage{physics}
\usepackage[normalem]{ulem}
\newcommand{\beq}{\begin{equation}}
\newcommand{\eeq}{\end{equation}}
\newcommand{\beqa}{\begin{eqnarray}}
\newcommand{\eeqa}{\end{eqnarray}}

\begin{document}

\preprint{APS/draft}

\title{Fast compression of pure-quartic solitons in nonlinear optical fibers via shortcuts to adiabaticity}

\author{Chengyu Han}
\affiliation{Department of Physics, Shanghai University, Shanghai 200444, China}

\author{Qian Kong}
\email{kongqian@shu.edu.cn}
\affiliation{Department of Physics, Shanghai University, Shanghai 200444, China}

\author{Ming Shen}
\email{shenmingluck@shu.edu.cn}
\affiliation{Department of Physics, Shanghai University, Shanghai 200444, China}

\author{Xi Chen}
\email{xi.chen@csic.es}
\affiliation{Instituto de Ciencia de Materiales de Madrid (CSIC), Cantoblanco, E-28049 Madrid, Spain}

\date{\today}

\begin{abstract}
Pure-quartic solitons (PQSs) supported by negative fourth-order dispersion have recently attracted considerable interest. In this work, we study both adiabatic and nonadiabatic compression of PQSs in nonlinear optical fibers with pure quartic dispersion in the presence of distributed gain and loss. Within a variational framework, we show that, for weak constant gain, the adiabatic compression dynamics can be mapped onto the motion of an effective particle in a slowly deformed potential, providing an intuitive physical picture. To overcome the long propagation distance required by conventional adiabatic condition, we exploit shortcuts to adiabaticity (STA) based on inverse engineering and derive analytical gain-loss profiles, with appropriate boundary conditions that realize a prescribed fast compression over a shorter propagation distance. Numerical simulations confirm the theoretical predictions and indicate a minimum propagation distance below which noticeable waveform distortion emerges. Compared with standard adiabatic references, the STA design significantly reduces the required compression distance while maintaining high-fidelity PQS evolution.
\end{abstract}

\maketitle
\section{introduction}

In nonlinear dispersive media such as optical fibers, temporal solitons arise from a balance between linear dispersion and nonlinear self-phase modulation \cite{agrawal}. Early studies focused predominantly on the standard regime governed by second-order (quadratic) dispersion, where soliton formation and manipulation have been extensively explored. Beyond this conventional setting, higher-order dispersion can qualitatively modify the pulse dynamics, enabling new propagation scenarios and affecting stability, spectral evolution, and waveform reshaping \cite{roy,liuwenjun,kruglov,rafat,qiang,fewo,hudson,karabo}.

Among higher-order effects, fourth-order (quartic) dispersion has attracted sustained attention over the past decades. Theoretically, solitons supported by dominant quartic dispersion have been predicted and analyzed in a variety of physical contexts, revealing distinct scaling laws and propagation characteristics compared with their quadratic-dispersion counterparts \cite{karlsson,karpman93,ivan,michel,karpman96,zambo,ziad,wamba,yulin,taki,bernd,sahoo,xutf}. More recently, the experimental observation of pure-quartic solitons (PQSs) enabled by negative pure quartic dispersion in silicon photonic-crystal waveguides \cite{nc} has spurred rapid progress in this area \cite{apl}, including studies of PQS generation and dynamics \cite{andrey,tam,yangchangxi,yao,justin,luozhichao,zhanglifu,tlidi,rivas,wabnitz,benjiamin}, pure-quartic soliton lasers \cite{darren}, self-similar propagation \cite{newton}, and modulation instability in quartic-dispersion systems \cite{conrad}.

A central goal in soliton physics and applications is pulse compression, which underpins high-peak-power pulse generation and nonlinear signal processing. Adiabatic compression—implemented through slow parameter variation to preserve soliton integrity—has been studied for decades \cite{Mollenauer,QuirogaTeixeiro,Harvey,Raju}, but it typically requires long propagation distances and can be sensitive to perturbations. Motivated by the development of shortcuts to adiabaticity (STA) \cite{muga}, inverse-engineering approaches have been proposed to accelerate soliton compression in matter waves and in optical settings while mitigating deleterious excitations \cite{lijing,chenxi,wangsiwei,2018Nonlinear,liyingjia,2020Manipulation,2020Shortcuts}. These works, however, have largely focused on quadratic dispersion/diffraction, whereas the role of dominant quartic dispersion in STA-enabled compression remains comparatively less addressed. Moreover, higher-order dispersion is known to introduce additional dynamical features—for instance, third-order dispersion can lead to temporal shifts of the soliton centroid during adiabatic evolution \cite{liyingjia}—highlighting the need for dedicated studies in quartic-dispersion soliton systems.

In this paper, we investigate adiabatic and nonadiabatic compression protocols for PQSs in nonlinear fibers with negative quartic dispersion in the presence of distributed gain and loss. Using a variational formulation, we derive reduced dynamical equations that provide a transparent effective-particle description of adiabatic PQS compression under weak constant gain. Inspired by this effective-particle picture, we apply STA based on inverse engineering to design gain–loss profiles that realize a prescribed fast compression over a shortened propagation distance, and we validate the theoretical design through direct numerical simulations. Our results demonstrate that STA protocols can substantially reduce the compression distance compared with standard adiabatic reference schemes while maintaining high-fidelity PQS evolution.

\section{Preliminaries}
\label{sec:preliminaries}

\subsection{Model and PQD-NLSE}
\label{subsec:model}

We consider temporal pulse propagation in a nonlinear optical fiber dominated by pure-quartic dispersion (PQD).
The slowly-varying envelope $\tilde\psi(T,Z)$ obeys the following nonlinear Schr\"{o}dinger equation (NLSE)
\cite{karlsson,karpman93,ivan,michel,karpman96,zambo,ziad,wamba,yulin,taki,bernd,sahoo,xutf}:
\begin{equation}
 i\frac{\partial \tilde{\psi}}{\partial Z}
+\frac{\tilde{\beta}_4}{24}\frac{\partial^4 \tilde{\psi}}{\partial T^4}
+\tilde{\gamma}|\tilde{\psi}|^2\tilde{\psi}
=i\tilde g(Z)\tilde{\psi},
\label{eq:NLSE_dim}
\end{equation}
where $Z$ is the propagation distance, $T$ is the retarded time, $\tilde{\beta}_4$ is the quartic-dispersion coefficient,
$\tilde{\gamma}$ is the Kerr coefficient, and $\tilde g(Z)$ describes distributed gain/loss.

To nondimensionalize the system, we introduce a reference duration $T_0$ and the quartic dispersion length
$L_DT_0^4/|\tilde{\beta}_4|$,
together with the normalized variables $t=T/T_0$ and $z=Z/L_D$, and the dimensionless gain $g(z)=\tilde g(Z)L_D$.
After normalization, the magnitude of $\tilde{\beta}_4$ is absorbed into $L_D$, while its \emph{sign} cannot be scaled out.
We therefore introduce
$ \beta_4=\tilde{\beta}_4/|\tilde{\beta}_4|=\pm 1 $,
so that the dimensionless dispersive term reads $(\beta_4/24)\,\partial_t^4\psi$ in Eq.~\eqref{eq:NLSE_nd}.
Furthermore, performing the gauge-amplitude transformation
\begin{equation}
\psi(t,z)=\tilde\psi(T,Z)\sqrt{\tilde\gamma L_D}\,
\exp\!\left(-\int_0^z g(z')\,dz'\right),
\label{eq:gauge}
\end{equation}
we remove the explicit gain/loss term and obtain the dimensionless PQD-NLSE with effective nonlinearity management,
\begin{equation}
 i\frac{\partial\psi}{\partial z}
+\frac{\beta_4}{24}\frac{\partial^4\psi}{\partial t^4}
+f(z)|\psi|^2\psi
=0,
\label{eq:NLSE_nd}
\end{equation}
where
\begin{equation}
 f(z)=\exp\!\left(2\int_0^z g(z')\,dz'\right).
\label{eq:fz_def}
\end{equation}
Physically, Eq.~\eqref{eq:NLSE_nd} shows that distributed gain/loss enters the conservative form of the model solely
through a longitudinal modulation of the effective Kerr strength $f(z)$.
Throughout this work we focus on the anomalous-PQD regime, $\beta_4=-1$ (equivalently $\tilde{\beta}_4<0$), so that
the dispersive term becomes $-(1/24)\partial_t^4\psi$.

Equation~\eqref{eq:NLSE_nd} conserves the $L^2$ norm (optical energy / power)
\begin{equation}
N=\int_{-\infty}^{\infty}|\psi(t,z)|^2\,dt,
\label{eq:power}
\end{equation}
which follows from global phase invariance (for real $f(z)$).
A natural Hamiltonian functional is
\begin{equation}
H(z)=\int_{-\infty}^{\infty}
\left[
\frac{1}{24}\left|\frac{\partial^2\psi}{\partial t^2}\right|^2
-\frac{f(z)}{2}|\psi|^4
\right]dt.
\label{eq:H_def}
\end{equation}
When $f(z)$ depends explicitly on $z$, the system is non-autonomous and $H(z)$ is \emph{not} conserved in general:
\begin{equation}
\frac{dH}{dz}
=-\frac{f'(z)}{2}\int_{-\infty}^{\infty}|\psi|^4\,dt.
\label{eq:H_balance}
\end{equation}
Hence $H$ is conserved only for constant $f$ (e.g., $g(z)\equiv 0$).

\subsection{Variational approximation with Gaussian ansatz}
\label{subsec:variational}

Equation~\eqref{eq:NLSE_nd} follows from the Lagrangian density
\begin{equation}
\mathcal{L}
=\frac{i}{2}\left(\psi^*\psi_z-\psi\psi_z^*\right)
-\frac{1}{24}\left|\psi_{tt}\right|^2
+\frac{f(z)}{2}|\psi|^4.
\label{eq:L_density}
\end{equation}
We employ a Gaussian trial function,
\begin{equation}
\psi(t,z)=A(z)\exp\!\left[-\frac{t^2}{2a^2(z)}+ic(z)t^2\right],
\label{eq:ansatz}
\end{equation}
where $A(z)$, $a(z)$, and $c(z)$ denote amplitude, temporal width, and quadratic chirp, respectively.
The conserved power \eqref{eq:power} gives
\begin{equation}
N=\int_{-\infty}^{\infty}|\psi|^2\,dt=\sqrt{\pi}\,A^2(z)a(z),
\label{eq:A_from_N}
\end{equation}
hence $A^2(z)=N/(\sqrt{\pi}\,a(z))$.
Substituting Eq.~\eqref{eq:ansatz} into Eq.~\eqref{eq:L_density} and integrating over $t$ yields the averaged Lagrangian
\begin{equation}
L
=\frac{N}{32a^4}
\left[
-16a^6\frac{dc}{dz}
-\left(4a^4c^2+1\right)^2
+8\sqrt{\frac{2}{\pi}}\,N\,a^3 f(z)
\right].
\label{eq:L_avg}
\end{equation}
The Euler--Lagrange equations read
\begin{align}
\frac{da}{dz}
&=\frac{c}{2a}\left(4a^4c^2+1\right),
\label{eq:ode_a}\\
\frac{dc}{dz}
&=-2a^2c^4
-\sqrt{\frac{2}{\pi}}\frac{N f(z)}{4a^3}
+\frac{1}{8a^6}.
\label{eq:ode_c}
\end{align}
These coupled first-order equations describe the variational evolution of the PQS width and chirp.
Differentiating Eq.~\eqref{eq:ode_a} with respect to $z$ and substituting Eqs.~\eqref{eq:ode_a}--\eqref{eq:ode_c}
eliminates $da/dz$ and $dc/dz$, yielding an exact second-order width equation
\begin{equation}
\frac{d^2 a}{dz^2}
=\frac{1}{16a^7}\left[
\left(1-\sqrt{\frac{8}{\pi}}\,N\,a^3 f(z)\right)
+\Delta_c(a,c;z)
\right],
\label{eq:dda_all}
\end{equation}
where the chirp-dependent correction is
\begin{equation}
\Delta_c(a,c;z)=16a^8c^4+8a^4c^2
-24\sqrt{\frac{2}{\pi}}\,N\,a^7c^2 f(z).
\label{eq:delta_c}
\end{equation}

In the adiabatic reference considered below, the evolution is slow and the chirp remains weak.
In particular, when $|4a^4c^2|\ll 1$, Eq.~\eqref{eq:ode_a} implies
\begin{equation}
\frac{da}{dz}\approx \frac{c}{2a}
\quad\Rightarrow\quad
c \sim 2a\,\frac{da}{dz},
\label{eq:c_scaling}
\end{equation}
so that $\Delta_c(a,c;z)=\mathcal{O}(c^2)+\mathcal{O}(c^4)$ represents higher-order corrections in the small adiabatic rate $da/dz$.
Neglecting $\Delta_c$ therefore yields the leading-order (Ermakov-like) width equation,
\begin{equation}
\frac{d^2a}{dz^2}
\approx \frac{1}{16a^7}
\left(
1-\sqrt{\frac{8}{\pi}}\,N\,a^3 f(z)
\right).
\label{eq:ermakov}
\end{equation}
This reduction reflects that the chirp mediates the energy exchange between PQD-induced dispersive ``pressure'' and Kerr
self-binding; under weak-chirp (adiabatic) evolution, the width follows an approximately autonomous dynamics.
However, for strongly nonadiabatic STA trajectories, the chirp-dependent contribution $\Delta_c(a,c;z)$ in
Eq.~\eqref{eq:dda_all} may become non-negligible. In that case, the inverse engineering should be
performed using the full variational system \eqref{eq:dda_all}, rather than the reduced weak-chirp
equation~\eqref{eq:ermakov}.

\section{Adiabatic reference}
\label{sec:adiabatic_reference}

 \begin{figure}[t]
	\centering
	\includegraphics[width=8.5cm]{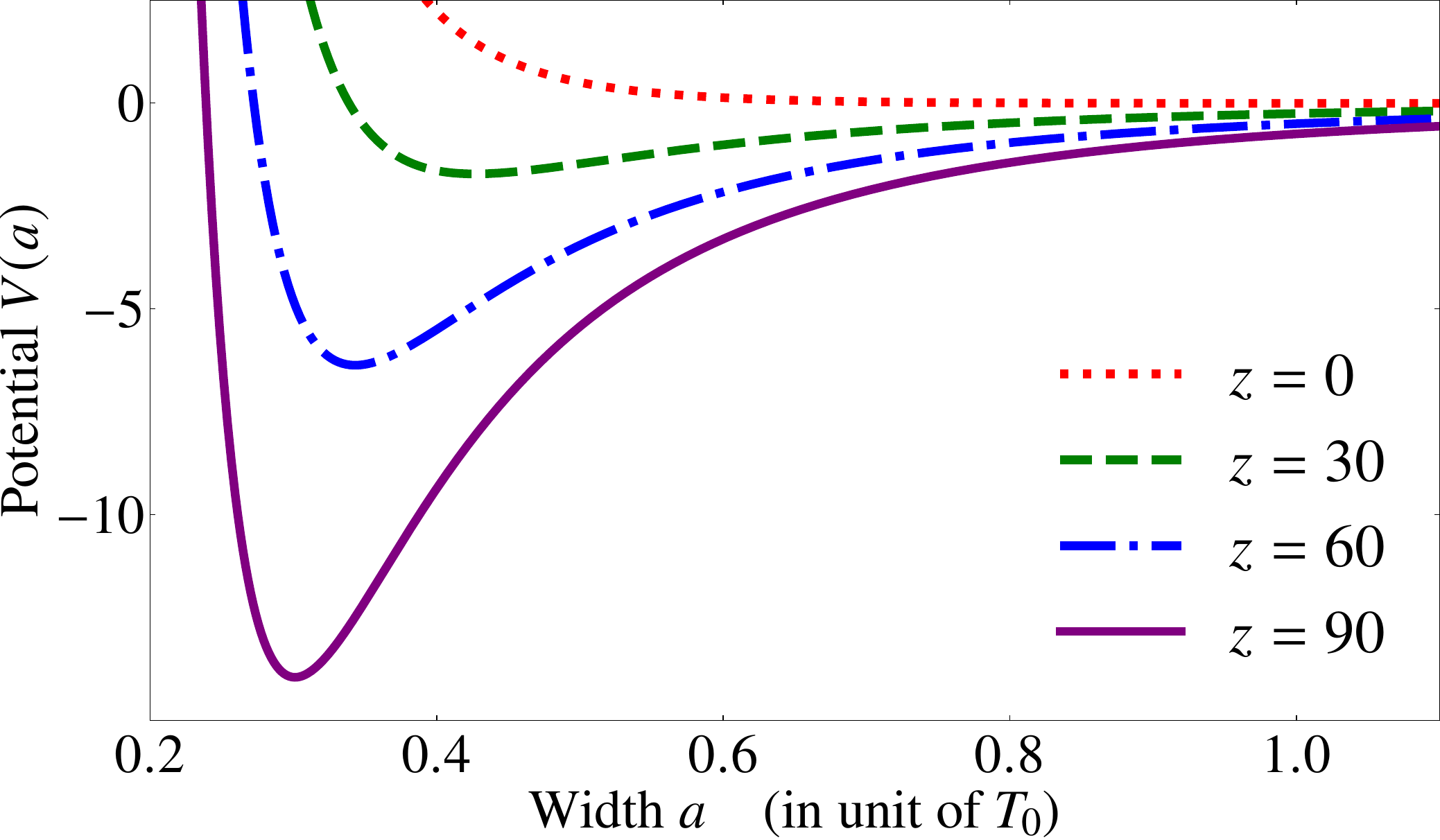}
	\caption{Effective potential $V(a)$ as a function of the PQS width $a$ at different propagation distances $z$ for constant gain $g(z)\equiv 0.02$ in the anomalous-PQD regime ($\beta_4=-1$). The curves correspond to $z=0$ (dotted), $z=30$ (dashed), $z=60$ (dash-dotted), and $z=90$ (solid). As $z$ increases, the minimum of $V(a)$ shifts to smaller $a$, indicating adiabatic compression driven by the gradual increase of the effective nonlinearity $f(z)=\exp(2gz)$.}
	\label{fig:Vaz}
\end{figure}

In this section, we shall establish an {adiabatic reference} for PQS compression, which will serve as the baseline for the STA
protocol discussed later. 
The starting point is the variational reduction \eqref{eq:ermakov}, by neglecting 
 the chirp-dependent correction $\Delta_c(a,c;z)$ in the adiabatic regime. In this scenario,
 a quasi-stationary (adiabatic) width then follows from the equilibrium condition $d^2a /dz^2=0$ in Eq.~\eqref{eq:ermakov},
\begin{equation}
a_c(z)=\left(\sqrt{\frac{\pi}{8}}\frac{1}{N f(z)}\right)^{1/3},
\label{eq:ac}
\end{equation}
which shows that distributed gain/loss affects the width only through the managed nonlinearity $f(z)$. If $f(z)$ varies slowly, one may freeze it locally and obtain a transparent mechanical interpretation: the width behaves as the
coordinate of an effective particle moving in an effective potential $V(a)$,
\begin{equation}
\left(\frac{da}{dz}\right)^2+V(a)=\mathcal{E},
\label{eq:energy_balance_z}
\end{equation}
with 
\begin{equation}
V(a)=\frac{1}{48a^6}-\frac{1}{24a^3}\sqrt{\frac{8}{\pi}}\,N\,f(z),
\label{eq:energy_balance}
\end{equation}
see Appendix~\ref{app:effective_potential}. The first term is the PQD-induced repulsive ``dispersive pressure'', while the second
term is the attractive Kerr binding strengthened by $f$. Notably, PQD yields a much more singular scaling than standard second-order group velocity dispersion (GVD) solitons \cite{liyingjia}:
the force contains an $a^{-7}$ contribution (equivalently $V\sim a^{-6}$), and the binding scales as $f\,a^{-3}$. Consequently, the
compression mechanism here is not a mere reparameterization of the usual GVD-soliton picture; instead, gain/loss acts primarily as
nonlinearity management that reshapes a PQD-specific potential landscape $V(a;z)\sim a^{-6}-f(z)a^{-3}$ and causes adiabatic drift
and/or weak breathing around its moving minimum.

\begin{figure}[t]
	\centering
	\includegraphics[width=8.5cm]{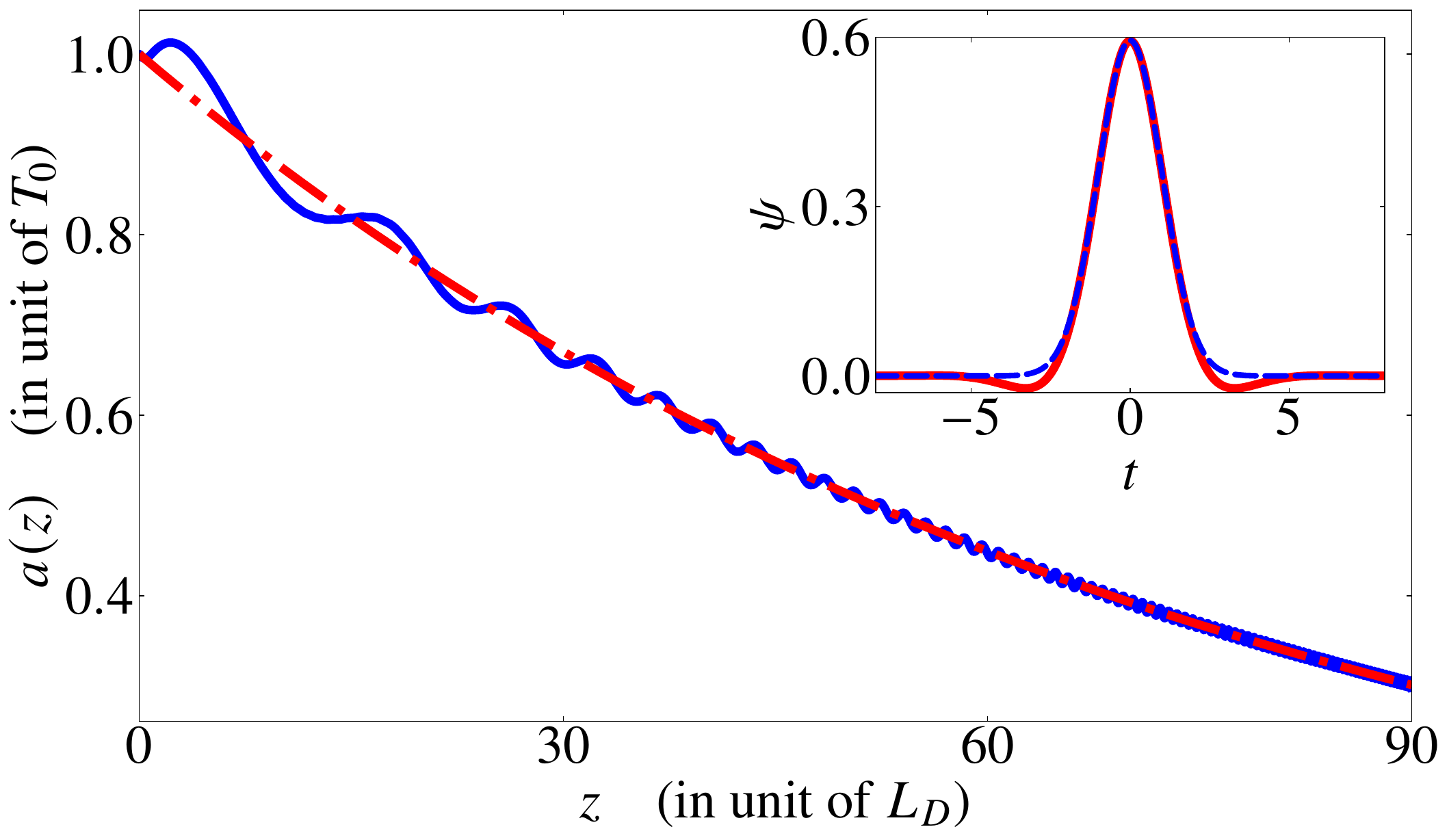}
\caption{Adiabatic evolution of the quasi-stationary width $a_c(z)$ versus propagation distance $z$. The dash-dotted curve shows the variational prediction from Eq.~\eqref{eq:ac}, whereas the solid curve shows the numerical simulation of Eq.~\eqref{eq:NLSE_nd}. The inset compares the initial Gaussian ansatz (dashed blue) with the numerically obtained PQS profile (solid purple). Parameters are consistent with Fig.~\ref{fig:Vaz}.}
	\label{fig:ac_num}
\end{figure}

It is instructive to classify the dynamics into three regimes, from the simplest autonomous case to the most general non-autonomous management.

\paragraph{Conservative case: $g(z)=0$, i.e., $f(z)=1$.}
The model becomes autonomous; both $N$ and $H$ are conserved. Stationary PQSs correspond to effective particles trapped at the minimum
of $V(a)$, and the stationary width (\ref{eq:ac}) is reduced to 
\begin{equation}
a_c^3=\frac{\sqrt{\pi/8}}{N}.
\end{equation}
Physically, PQD-induced dispersive pressure and Kerr self-binding exactly balance at the potential minimum, yielding a robust PQS.

\paragraph{Constant gain: $g(z)=g_0>0$, i.e., $f(z)=e^{2g_0 z}$.}
In the gauge-transformed conservative form Eq. \eqref{eq:NLSE_nd}, the power $N$ remains conserved, whereas the effective nonlinearity
$f(z)$ grows monotonically and continuously reshapes the potential $V(a)$. As a result, the potential minimum drifts toward smaller
$a$, leading to adiabatic compression. This behavior is illustrated in Fig.~\ref{fig:Vaz}, where increasing $z$ shifts the minimum of
$V(a)$ to smaller widths. The quasi-stationary width is given by Eq.~\eqref{eq:ac}, and for constant gain it yields
\begin{equation}
a_c(z)=a_c(0)\,e^{-\frac{2}{3}g_0 z},
\end{equation}
which provides a benchmark distance to reach a target width $a_f$,
\begin{equation}
z_f^{(\mathrm{adiab})}=\frac{3}{2g_0}\ln\!\left(\frac{a_c(0)}{a_c(z_f)}\right).
\label{eq:adiabatic_distance}
\end{equation}
For $a_c(0)=1$, $a_c(z_f)=0.3$, and $g_0=0.02$, one obtains $z_f^{(\mathrm{adiab})}\simeq 90$, consistent with the numerical adiabatic
trajectory shown in Fig.~\ref{fig:ac_num}. In this case, although the Gaussian ans\"atz serves as an approximate description of the PQS, it effectively captures the core features of the PQS dynamical evolution \cite{evolution}.

\paragraph{General distributed gain/loss: $g(z)$ , i.e., nonlinearity management.}
An arbitrary profile $g(z)$ implements general nonlinearity management through $f(z)$ in Eq.~\eqref{eq:fz_def}, see also Appendix \ref{app:adiabaticity_criterion}.
If $g(z)$ is sufficiently weak and/or varies slowly, the evolution is adiabatic and the width follows the drifting equilibrium
$a_c(z)$ in Eq.~\eqref{eq:ac}. By contrast, a rapidly varying $g(z)$ drives the system out of the adiabatic manifold, resulting in
nonadiabatic excitations and pronounced breathing around the moving minimum. This separation into adiabatic and nonadiabatic regimes
provides the physical motivation for STA: one aims to reach the same final width over a much shorter distance while controlling the
excess breathing.

Finally, we emphasize that the weak-chirp reduction is tailored to the adiabatic reference: for strongly nonadiabatic STA trajectories
the chirp-dependent correction may become non-negligible, and inverse engineering should then be carried out using the full variational
system \eqref{eq:ode_a}--\eqref{eq:ode_c}. This motivates the STA design introduced in the next section.

\section{Inverse-Engineering for STA}
\label{sec:sta}

Adiabatic compression of PQSs can be achieved by using a weak (approximately constant) gain coefficient $g$, which implies a slowly varying effective nonlinearity $f(z)=\exp\!\big(2\int_0^z g(z')dz'\big)$. However, the adiabatic requirement necessarily leads to a long compression distance. To realize fast and robust compression over a much shorter distance, we employ a STA strategy based on inverse engineering \cite{lijing,liyingjia}. The key idea is to prescribe a smooth width trajectory $a(z)$ that connects the same initial and final widths as the adiabatic reference, and then reconstruct the required gain/loss profile $g(z)$ from the reduced width dynamics.

We impose the STA boundary conditions such that the compressed state matches the adiabatic reference at the beginning and at the end of the process,
\begin{equation}
a(0)=a_c(0),\qquad a(z_f)=a_c(z_f),
\label{eq:sta_bc1}
\end{equation}
where $z_f$ denotes the designed compression distance. In our adiabatic reference, $z_f=90$, whereas in STA we target much smaller values of $z_f$ while keeping the same compression ratio. In addition, to suppress spurious excitations and guarantee a smooth connection to stationary PQSs at $z=0$ and $z=z_f$, we require the first few derivatives of the width to vanish at both ends \cite{lijing},
\begin{equation}
\frac{d^k a}{dz}(0)=\frac{d^k a}{dz}(z_f)=0,\qquad k=1,2,3.
\label{eq:sta_bc2}
\end{equation}
These smoothness conditions ensure that the auxiliary degrees of freedom (notably the chirp) are not abruptly excited at the endpoints and that the STA trajectory is compatible with a quasi-stationary soliton at both boundaries.
A convenient interpolation that satisfies Eqs.~\eqref{eq:sta_bc1}--\eqref{eq:sta_bc2} is a seventh-order polynomial ans\"atz,
\begin{equation}
a(z)=\sum_{j=0}^{7} a_j z^j,
\label{eq:poly}
\end{equation}
where the coefficients $a_j$ are uniquely determined by the eight boundary conditions in Eqs.~\eqref{eq:sta_bc1}--\eqref{eq:sta_bc2}.
Once $a(z)$ is fixed, the effective nonlinearity $f(z)$ can be determined numerically via Eqs. (\ref{eq:ode_a}) and (\ref{eq:ode_c}).
Finally, the gain/loss profile $g(z)$ is obtained from the definition $f(z)=\exp\!\big(2\int_0^z g(z')dz'\big)$,
\begin{equation}
g(z)=\frac{1}{2}\frac{d}{dz} \left[\ln f(z)\right],
\label{eq:g_inverse}
\end{equation}
which constitutes the core of the inverse-engineering STA protocol: the desired width trajectory uniquely determines $f(z)$ and hence $g(z)$.

\begin{figure}[t]
	 \centering\includegraphics[width=9cm]{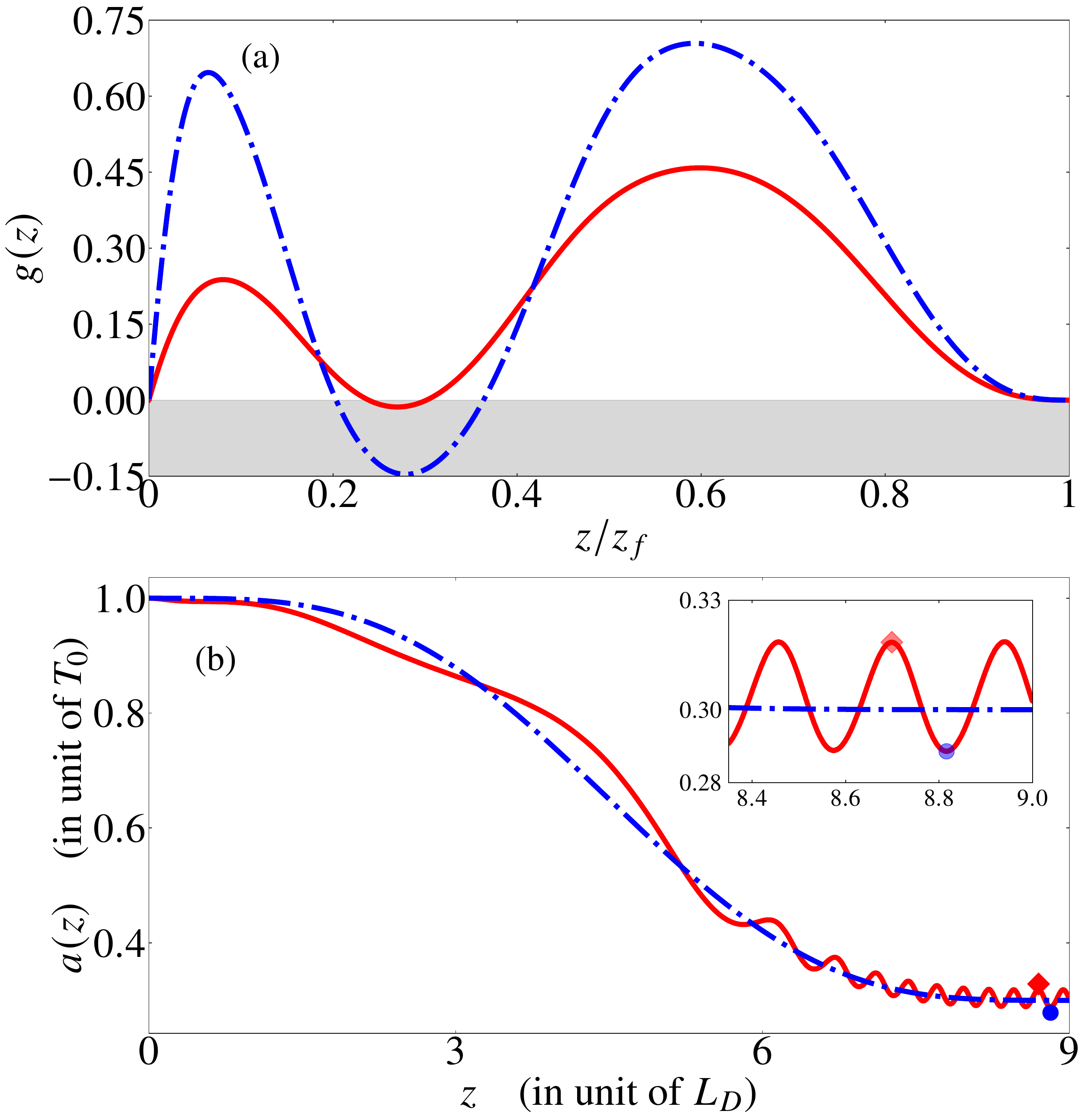}
	\caption{(a) STA-designed gain/loss profiles $g(z)$ for two compression distances, $z_f=9$ (solid red) and $z_f=6$ (dash-dotted blue), under the same boundary conditions $a(0)=1$ and $a(z_f)=0.3$. (b) Evolution of the width $a$ of the pulse versus propagation distance $z$ at $z_f=9$ . The solid red and dash-dotted blue curves represent direct numerical calculation and the STA-designed trajectory, respectively.}
	\label{fig:sta_two_panels}
\end{figure}

Figure~\ref{fig:sta_two_panels} (a) shows two examples of STA-designed $g(z)$ for compression distances $z_f=9$ and $z_f=6$ under identical boundary conditions $a(0)=1$ and $a(z_f)=0.3$. As expected, reducing $z_f$ requires a stronger modulation amplitude (larger peak gain). Interestingly, the optimal profile generally involves not only gain but also a loss segment (i.e., $g(z)<0$ over a finite interval). Physically, this ``gain-loss'' sequence accelerates the compression initially and then damps the induced breathing so that the trajectory arrives at $a(z_f)$ with a small residual velocity and curvature, consistent with the endpoint conditions in Eq.~\eqref{eq:sta_bc2}.


\begin{figure}[ht!]
	 \centering\includegraphics[width=7cm]{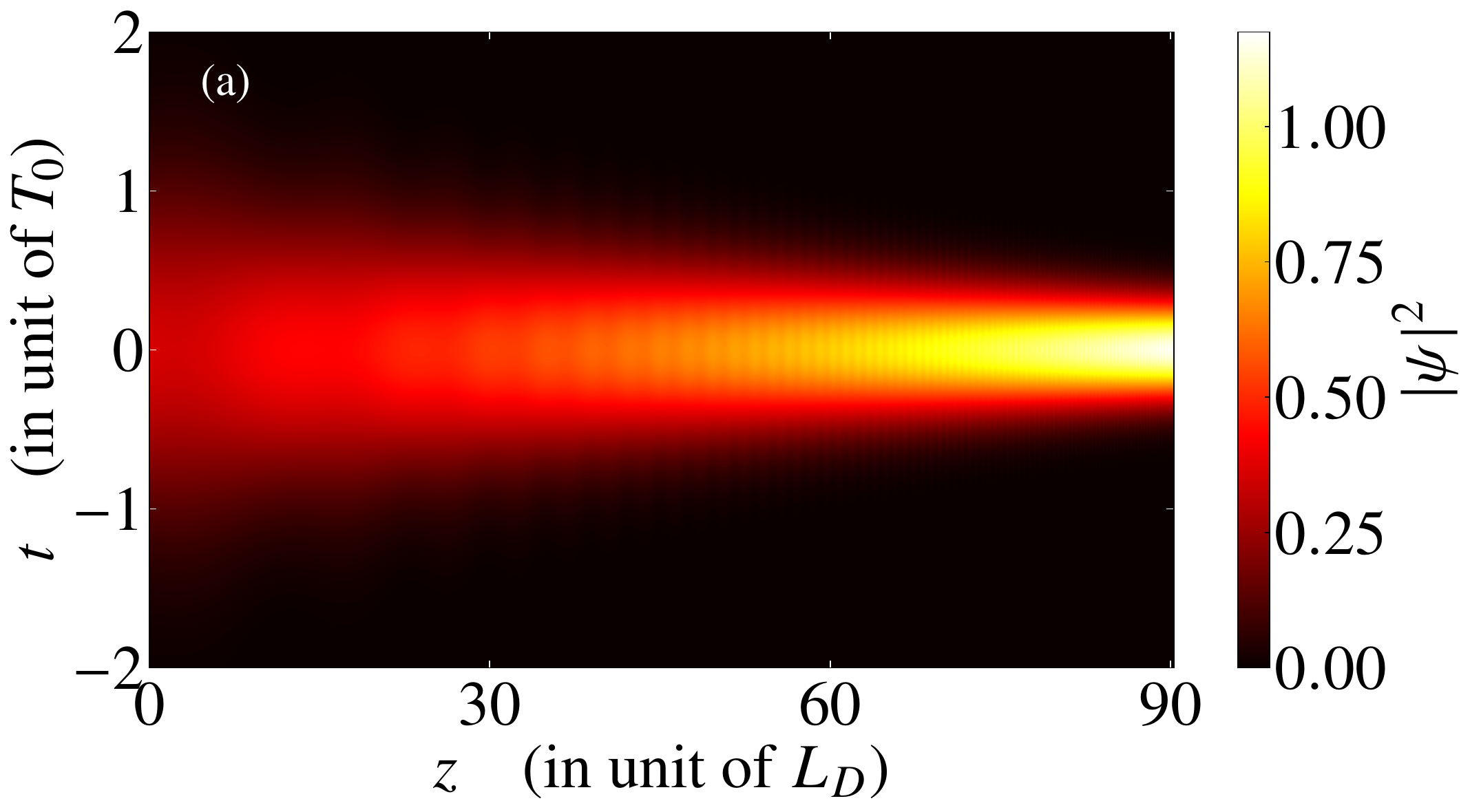}
 \centering\includegraphics[width=7cm]{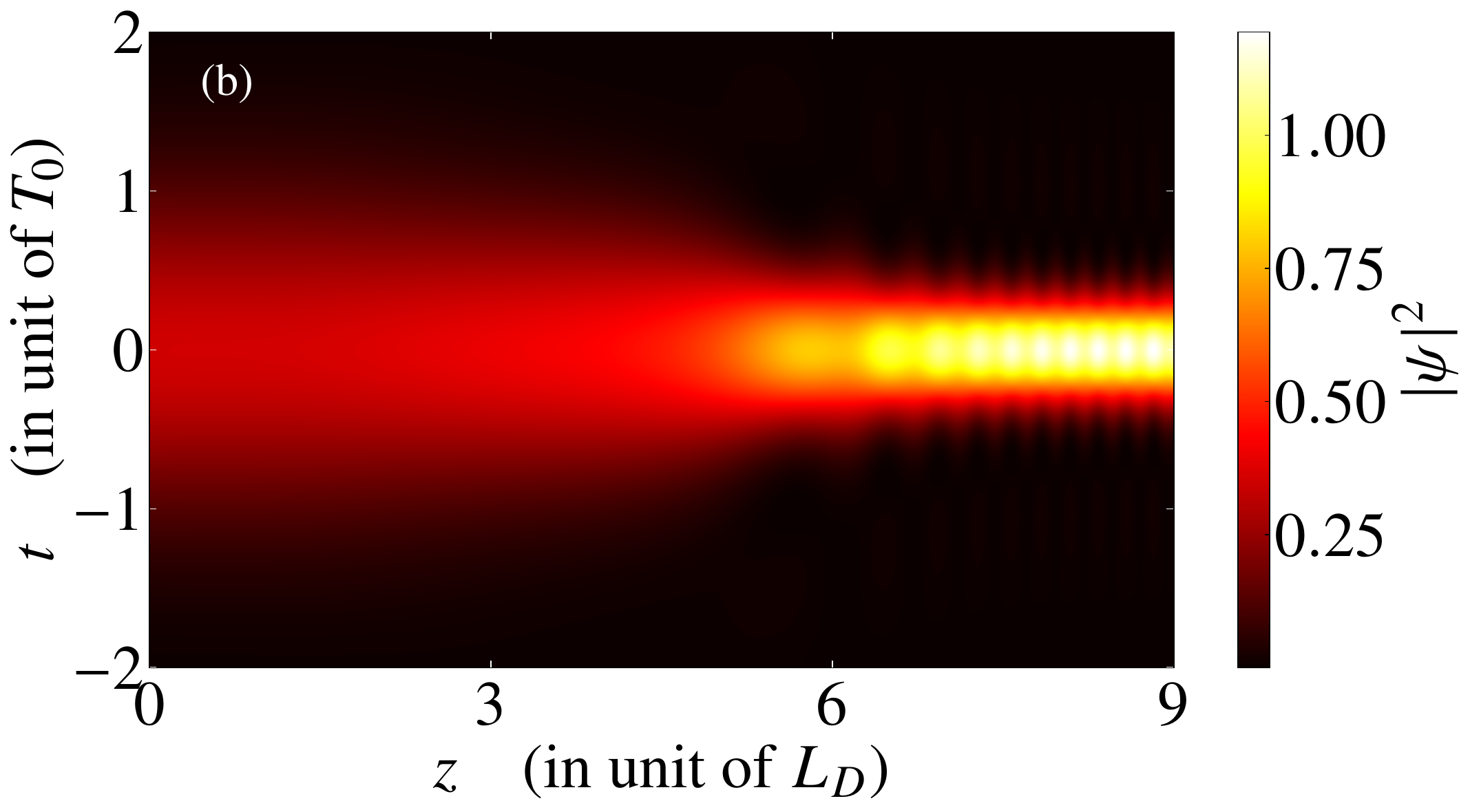}
	\centering\includegraphics[width=4cm]{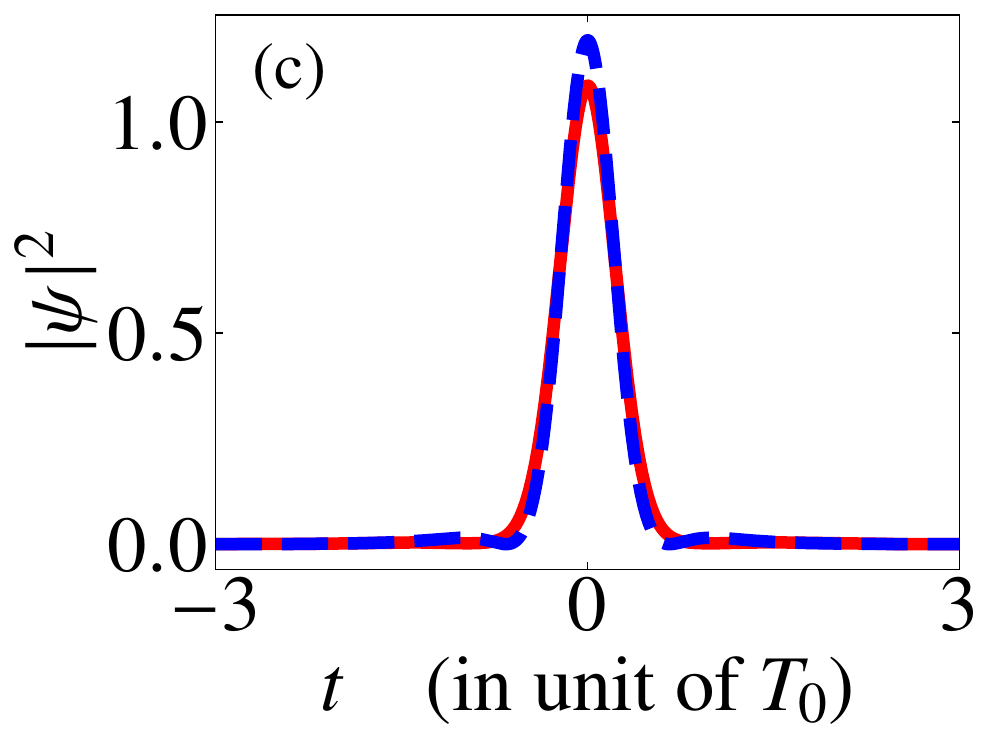}
	\centering\includegraphics[width=4cm]{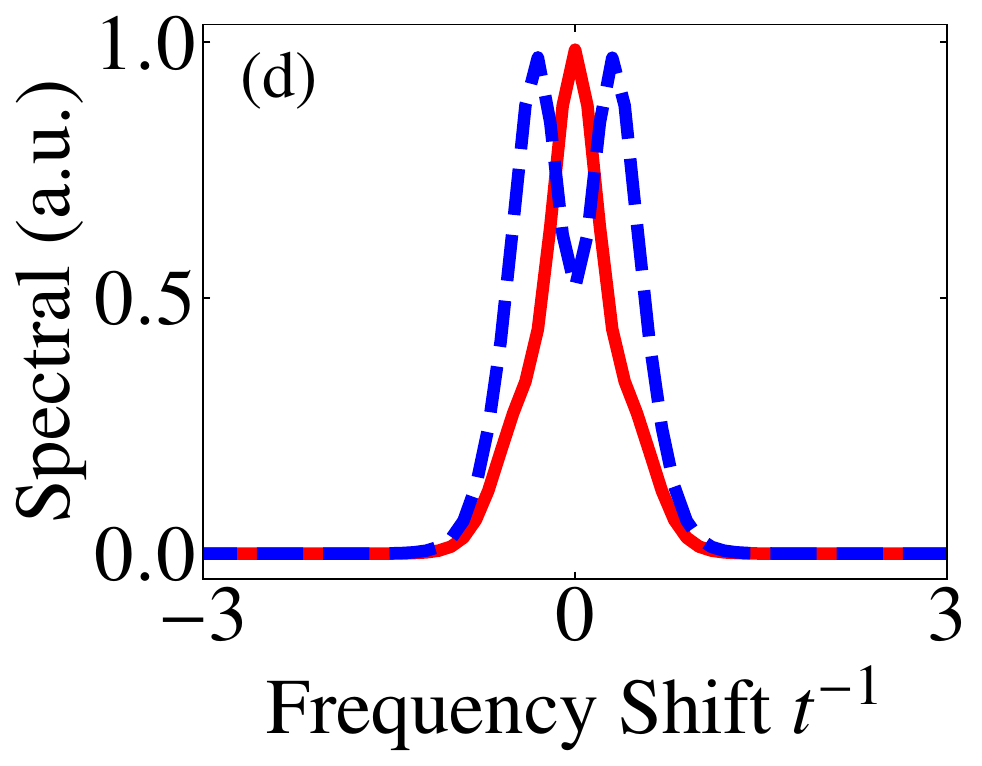}
	\caption{Spatiotemporal evolution of Gaussian soliton under the adiabatic (a) and STA (b) protocols, respectively. Parameters are consistent with Figs.~\ref{fig:ac_num} and ~\ref{fig:sta_two_panels}. Panels (c) and (d) show the intensity profiles and normalized spectra at the
instants when the width $a(z)$ is maximal (solid red) and minimal (dashed blue).}
\label{fig:sta_spacetime}
\end{figure}

Using the STA-designed $g(z)$, Fig.~\ref{fig:sta_two_panels} (b) compares the width evolution $a(z)$ predicted by the STA design (dash-dotted blue line) with the direct numerical simulation of Eq.~\eqref{eq:dda_all} (solid red line). Excellent agreement is observed, demonstrating that the PQS can be compressed to $a(z_f)=0.3$
within a remarkably short distance (e.g.\ $z_f=9$), i.e.\ roughly an order of magnitude shorter than the adiabatic reference.
Unlike adiabatic compression, where the evolution follows the slowly drifting equilibrium $a_c(z)$ determined by the instantaneous
fixed point, the STA protocol exploits the full dynamical law for $a(z)$ (including acceleration and deceleration) to steer the system between the same endpoints while suppressing excess excitations through the boundary conditions.

The spatiotemporal evolution under the adiabatic and STA protocols is shown in Fig.~\ref{fig:sta_spacetime}(a) and (b).
Near the final distance $z_f$, the numerical width exhibits weak residual breathing around $a(z_f)=0.3$.
This behavior can be attributed to (i) unavoidable excitation of internal modes during a strongly nonadiabatic process, and
(ii) the approximate nature of the Gaussian ansatz used in the reduced model, which captures the PQS core but does not fully reproduce
the weak oscillatory tail of the exact PQS \cite{evolution}. Additional deviations may also arise from dispersive radiation neglected in
the reduced description \cite{evolution,tsoy2024generic}. Consequently, an STA protocol designed within the reduced (Gaussian) manifold
can leave a small mismatch, see also the inset in Fig. \ref{fig:sta_two_panels}(b), when implemented in the full field dynamics.

To characterize the breathing cycle, Figs.~\ref{fig:sta_spacetime}(c) and \ref{fig:sta_spacetime}(d) show the intensity profiles and
normalized spectra at the instants when the width is maximal and minimal, respectively.
At the strongest compression point, the numerical pulse may develop a weak pedestal, indicating that the nonlinear phase modulation
temporarily exceeds that of an exact fundamental soliton and induces an additional nonlinear chirp; subsequent dispersion converts this
chirp into a pedestal-like structure, analogous to features observed in higher-order soliton compression \cite{Takashi}. In the spectral
domain, the enhanced nonlinear chirp manifests as spectral distortion (including a tendency toward multi-peaked spectra) \cite{Takashi}.
Driven by the interplay between PQD and Kerr nonlinearity, this excitation settles into a stable oscillation, consistent with the
width oscillations observed after the compression stage.
Notably, the observed breathing period ($0.234$) is close to the internal-mode period ($0.233$) reported in Ref.~\cite{tam}, suggesting
that the post-compression dynamics can be interpreted as a weakly excited internal mode of the PQS. This also indicates that, even after fast STA compression, the PQS may sustain long-range stable breathing around the target width $a(z_f)$.

\begin{figure}[t]
	\centering\includegraphics[width=8.2cm]{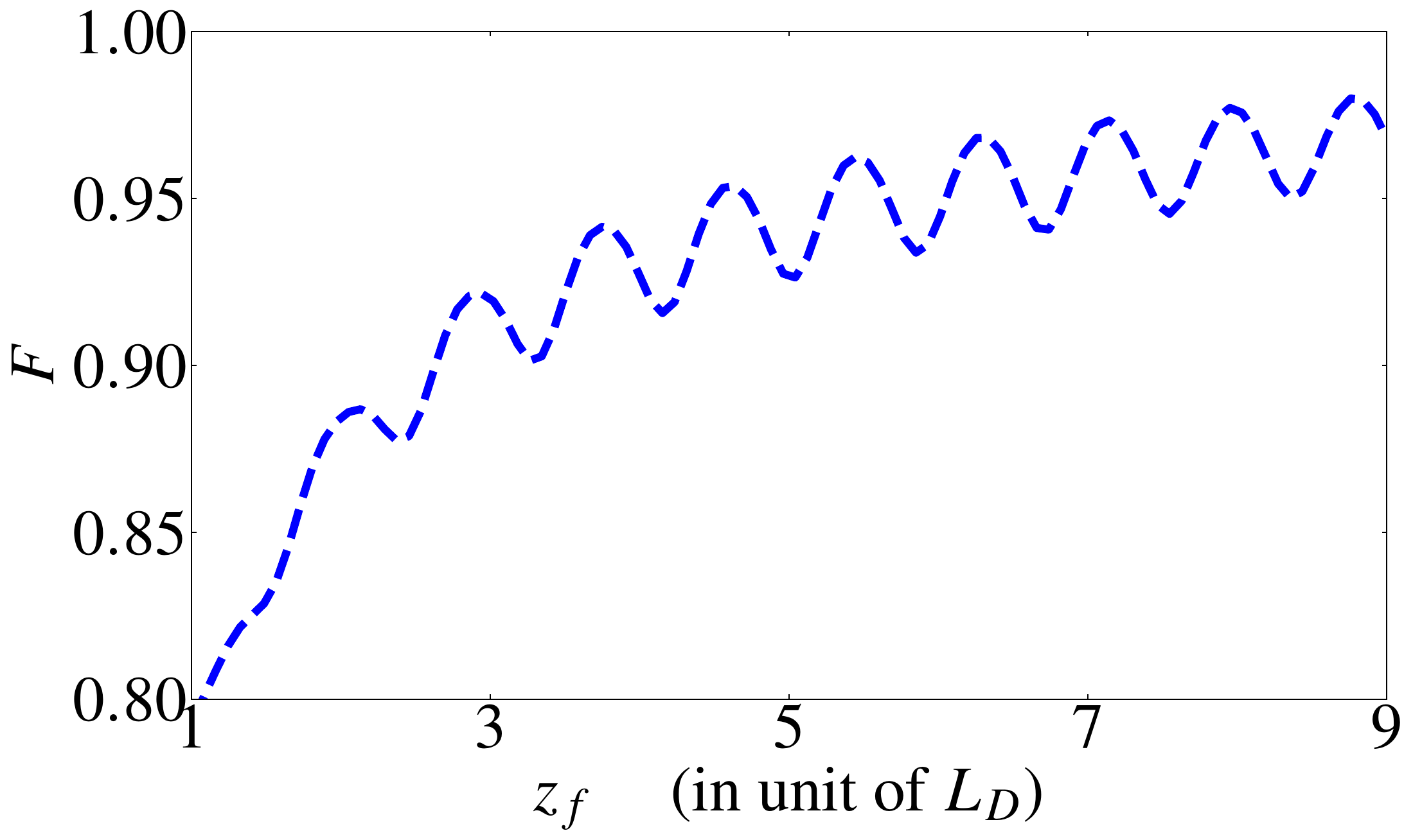}
	\caption{Fidelity $F$ as a function of propagation distance $z_f$ for STA protocols. Parameters are the same as those in Fig. \ref{fig:sta_two_panels}.}
    \label{fig:fidelity}
\end{figure}

To evaluate the performance of STA at reduced compression distances, we quantify the overlap between the numerically obtained field
$\tilde{\psi}(z_f,t)$ and the variational ansatz $\psi(z_f,t)$ [Eq.~\eqref{eq:ansatz}] by the fidelity
\begin{equation}
F=\big|\braket{\psi(z_f,t)}{\tilde{\psi}(z_f,t)}\big|^{2},
\end{equation}
where both fields are normalized to unity. Note that for each chosen compression distance $z_f$, the corresponding gain/loss profile
$g(z)$ (and hence $f(z)$) must be designed separately by the inverse-engineering procedure.

As shown in Fig.~\ref{fig:fidelity}, the STA protocol (e.g., $z_f\ge 9$) achieves the prescribed compression over a propagation
distance that is nearly an order of magnitude shorter than the adiabatic reference ($g_0=0.02$, $z_f=90$). Near the STA endpoint, $F$
exhibits weak oscillations as a function of $z_f$, which originate from residual breathing of the compressed PQS, as discussed above.
In general, one may further reduce $z_f$ by increasing the modulation strength of the gain/loss profile. For instance, $z_f=6$
provides a practical compromise between compression efficiency and waveform quality, for which the fidelity remains high
($F\simeq 0.968$). Importantly, however, the compression distance cannot be reduced arbitrarily: shorter $z_f$ requires larger peak
gain/loss, which enhances intermediate excitations (e.g., breathing and dispersive radiation) and increases the mismatch between the
reduced variational manifold and the full field dynamics. As a result, the STA protocol constructed from the variational
approximation eventually loses accuracy. A complementary theoretical estimate of a gain/loss--limited lower bound on $z_f$ is provided
in Appendix~\ref{sec:mvt_bound}.

\section{Conclusion and outlook}
\label{sec:conclusion}

In conclusion, we have investigated both adiabatic and nonadiabatic compression of pure-quartic solitons (PQSs) in nonlinear optical
fibers with distributed gain/loss. For weak constant gain, we established an adiabatic reference within a Gaussian variational
approximation, leading to an effective-potential picture in which the soliton width follows the slowly drifting minimum of a gradually
reshaped potential. Building on this reference, we implemented a shortcut-to-adiabaticity (STA) protocol via inverse engineering: by
prescribing a smooth width trajectory subject to appropriate endpoint conditions, we reconstructed the required longitudinal gain/loss
profile $g(z)$ (equivalently, the managed nonlinearity $f(z)$). The resulting STA design enables rapid nonadiabatic compression to the
same target width while reducing the required propagation distance by nearly one order of magnitude (from $z_f\simeq 90$ to $z_f\simeq 9$ for the
same compression ratio). Direct
numerical simulations of the governing equation confirm the effectiveness of the STA approach and reveal weak residual breathing near
the target distance, consistent with excitation of PQS internal modes and with the approximate nature of the Gaussian ansatz.

Several extensions are natural. First, the STA design can be further optimized by employing more accurate trial functions (or multi-parameter ans\"atze) \cite{Qientropy} or even Pontryagin's maximum principle \cite{Tangyoupra} that capture the oscillatory tails of PQSs, thereby reducing residual breathing and pedestal formation. Second, robustness against experimental imperfections---such as bounded peak gain, noise, and parameter uncertainty in $\beta_4$ and $\gamma$---can be incorporated by constrained or robust inverse engineering. Third, it would be of interest to generalize the method to jointly managed dispersion and nonlinearity, as well as to include higher-order physical effects (e.g., self-steepening or Raman response) to assess the practical performance of fast PQS compression in realistic platforms. These directions would strengthen the connection between STA control concepts and engineered ultrafast pulse shaping in PQD-based nonlinear waveguides.

\section*{Acknowledgments}
This work is supported by project Grant No. PID2021-125823NA-I00 funded by MCIN/AEI/10.13039/501100011033 and by ``ERDF A way of making Europe" and ``ERDF Invest in your Future'',  Basque Government through Grant No. IT1470-22, through the ELKARTEK program, project KUBIT (KK-2024/00105), programa CSIC-NSTC (BINST24012), and the Severo Ochoa Centres of Excellence program through Grant CEX2024-001445-S. 

\appendix
\section{Effective-potential formulation}
\label{app:effective_potential}

In this appendix we clarify what the effective potential is, how it is derived from the variational width equation, and under which conditions it provides a valid physical picture. Starting from the Gaussian variational reduction, the pulse width $a(z)$ obeys the Ermakov-like equation
\begin{equation}
\frac{d^2 a}{d z^2}
=\frac{1}{16a^{7}}
\left(
1-\sqrt{\frac{8}{\pi}}\,N\,a^3 f(z)
\right),
\label{eq:app_width}
\end{equation}
where $N=\int_{-\infty}^{\infty}|\psi|^2\,dt$ is the conserved power of Eq.~\eqref{eq:NLSE_nd} and $f(z)=\exp\!\left(2\int_0^z g(z')dz'\right)$ is the effective nonlinearity induced by distributed gain/loss. When $f(z)$ varies sufficiently slowly compared with the characteristic width-oscillation length scale, one may adopt a {local} (adiabatic) approximation in which $f(z)$ is treated as a constant parameter within a short propagation window. We denote this locally frozen value by $f$ and introduce
\begin{equation}
\kappa=\sqrt{\frac{8}{\pi}}\,N\,f \qquad \text{(locally constant)}.
\label{eq:app_kappa}
\end{equation}
Then Eq.~\eqref{eq:app_width} becomes an autonomous Newton-type equation
\begin{equation}
\frac{d^2 a}{dz^2}=\frac{1}{16a^7}-\frac{\kappa}{16a^4}.
\label{eq:app_width_autonomous}
\end{equation}

For an autonomous equation of the form $d^2 a/dz=F(a)$ with the unit mass of fictitious particle, one may define an effective potential $V(a)$ by
\begin{equation}
\frac{d^2 a}{dz^2}=-\frac{1}{2}\frac{dV}{da},
\label{eq:app_V_def}
\end{equation}
which implies the energy-balance (first integral)
\begin{equation}
\left(\frac{da}{dz}\right)^2 + V(a)=\mathcal{E},
\label{eq:app_energy_balance}
\end{equation}
with $\mathcal{E}$ an integration constant determined by initial conditions.
Using Eq.~\eqref{eq:app_width_autonomous} in Eq.~\eqref{eq:app_V_def}, we obtain
\begin{equation}
\frac{dV}{da}=-\left(\frac{1}{8a^7}-\frac{\kappa}{8a^4}\right).
\label{eq:app_V_prime}
\end{equation}
Integrating with respect to $a$ yields (up to an irrelevant additive constant)
\begin{equation}
V(a)=\frac{1}{48a^6}-\frac{\kappa}{24a^3}
=\frac{1}{48a^6}-\frac{1}{24a^3}\sqrt{\frac{8}{\pi}}\,N\,f.
\label{eq:app_V}
\end{equation}
Eq.~\eqref{eq:app_V} contains two competing contributions:
(i) the quartic-dispersion term $V_D(a)=1/(48a^6)$, which is strongly repulsive for narrow pulses ($a\to 0$), reflecting the rapid increase of dispersive ``cost'' under PQD;
(ii) the nonlinear term $V_{NL}(a)=-(\kappa)/(24a^3)$, which is attractive and represents Kerr self-binding enhanced by $f$.
A stable quasi-stationary width corresponds to a minimum of $V(a)$, i.e. $dV/da=0$, which reproduces the equilibrium condition from $d^2 a/dz=0$.

If $f(z)$ depends explicitly on $z$, the system is non-autonomous and the quantity $\mathcal{E}$ in Eq.~\eqref{eq:app_energy_balance} is generally not conserved. In that case the potential becomes slowly $z$-dependent, $V(a)$, and Eq.~\eqref{eq:app_energy_balance} should be regarded as a \emph{local} diagnostic rather than an exact invariant.
The effective-potential picture remains useful in the adiabatic regime where the modulation length of $f(z)$ is much larger than the intrinsic width-oscillation length, so that $f(z)$ changes little over one oscillation cycle. Physically, a slow increase of $f(z)$ deepens the attractive part of $V(a)$ and shifts its minimum toward smaller $a$, providing an intuitive explanation for adiabatic compression and/or weak breathing of PQD solitons under distributed gain.

\section{Adiabatic criterion}
\label{app:adiabaticity_criterion}

A practical criterion for adiabatic following can be obtained by comparing the relative modulation rate of $f(z)$ with the intrinsic
small-oscillation frequency of the width around the instantaneous equilibrium. Using the locally frozen effective potential
$V(a)$ [Eq.~\eqref{eq:app_V}], we expand $V$ to second order around $a=a_c$ and obtain the harmonic approximation
\begin{equation}
\delta a'' + \Omega_a^2\,\delta a \approx 0,\qquad \delta a \equiv a-a_c,
\end{equation}
with
\begin{equation}
\Omega_a^2=\frac{1}{2}V''(a_c)=\frac{3}{16}\,a_c^{-8},
\qquad
\Omega_a=\frac{\sqrt{3}}{4}\,a_c^{-4}.
\label{eq:Omega_a}
\end{equation}
A convenient sufficient condition for adiabatic following is therefore
\begin{equation}
\left|\frac{f'(z)}{f(z)}\right|\ll \Omega_a(z),
\label{eq:adiabatic_condition}
\end{equation}
i.e., $f(z)$ should not change appreciably over one width-oscillation period. For constant gain $g(z)=g_0$, one has $f'/f=2g_0$ and
$a_c(z)\propto f(z)^{-1/3}$, hence $\Omega_a(z)\propto a_c^{-4}\propto f(z)^{4/3}=\exp(8g_0 z/3)$. Therefore, if the adiabatic
criterion is satisfied initially, it becomes progressively better satisfied as $z$ increases.

When $f$ depends explicitly on $z$, the mechanical ``energy'' $\mathcal{E}$ in Eq.~\eqref{eq:app_energy_balance} is no longer an exact
invariant because the potential is explicitly $z$-dependent. Differentiating Eq.~\eqref{eq:app_energy_balance} gives the energy drift
\begin{equation}
\frac{d\mathcal{E}}{dz}
=\frac{\partial V}{\partial z}
=-\frac{1}{24a^3}\sqrt{\frac{8}{\pi}}\,N\,f'(z).
\label{eq:energy_drift}
\end{equation}
In the adiabatic regime where $a(z)\approx a_c(z)$, Eq.~\eqref{eq:energy_drift} explains why a slow increase of $f(z)$ primarily
induces a gradual shift of the equilibrium (compression), whereas a faster variation of $f(z)$ injects energy into the width degree of
freedom and manifests as noticeable breathing. This provides a quantitative baseline for STA: the goal of inverse engineering is to
drive $a(z)$ from $(a(0),a'(0))$ to $(a(z_f),a'(z_f))$ over a short distance $z_f$ while minimizing the excess breathing associated with
nonadiabatic energy injection. Here and in the following, a prime denotes differentiation with respect to the propagation distance $z$, i.e.,
$a'(z)\equiv da/dz$ and $a''(z)\equiv d^2a/dz^2$.

\section{Bound for propagation distance}
\label{sec:mvt_bound}

In the adiabatic regime, the reduced width dynamics admits an energy-balance form
\begin{equation}
\left(\frac{da}{dz}\right)^2+V(a)\approx \mathcal{E},
\label{eq:adiabatic_energy_balance}
\end{equation}
which leads to the time-of-flight (distance) estimate
\begin{equation}
z_f \approx \int_{a_f}^{a_i}\frac{da}{\sqrt{\mathcal{E}-V(a)}}.
\label{eq:zf_energy_quadrature}
\end{equation}
For general STA trajectories, however, $f(z)$ varies appreciably and the width dynamics is non-autonomous; therefore
Eqs.~\eqref{eq:adiabatic_energy_balance}--\eqref{eq:zf_energy_quadrature} do \emph{not} define a global invariant and cannot by
themselves provide a rigorous lower bound on $z_f$.

A robust and practical lower bound can instead be derived directly from the nonlinearity-management law,
\begin{equation}
f(z)=\exp\!\left(2\int_{0}^{z} g(s)\,ds\right),
\label{eq:f_management}
\end{equation}
Assume that the STA protocol connects two quasi-stationary PQSs (no breathing at the endpoints), as enforced by the boundary
conditions~\eqref{eq:sta_bc1}--\eqref{eq:sta_bc2}. Then the fixed-point relation of the reduced width dynamics implies
\begin{equation}
a_c^3(z)=\frac{\sqrt{\pi/8}}{N f(z)}
\quad\Longrightarrow\quad
\frac{f(z_f)}{f(0)}=\left(\frac{a_c(0)}{a_c(z_f)}\right)^3,
\label{eq:f_ratio_from_endpoints}
\end{equation}
where $a_f=a_c(z_f)$ is the target width. Combining Eq.~\eqref{eq:f_management} at $z=z_f$ with
Eq.~\eqref{eq:f_ratio_from_endpoints} gives
\begin{equation}
\ln\!\left(\frac{f(z_f)}{f(0)}\right)=2\int_{0}^{z_f} g(s)\,ds
=3\ln\!\left(\frac{a_c(0)}{a_c(z_f)}\right).
\label{eq:logf_identity}
\end{equation}

Now apply the mean-value theorem for integrals: since $g$ is integrable on $[0,z_f]$, there exists $\xi\in(0,z_f)$ such that
\begin{equation}
\int_{0}^{z_f} g(s)\,ds = g(\xi)\,z_f,
\label{eq:MVT_integral}
\end{equation}
and hence
\begin{equation}
\ln\!\left(\frac{f(z_f)}{f(0)}\right)=2g(\xi)\,z_f.
\label{eq:logf_MVT}
\end{equation}
If the gain/loss amplitude is physically bounded as $|g(z)|\le g_{\max}$ on $[0,z_f]$, then $|g(\xi)|\le g_{\max}$ and
Eq.~\eqref{eq:logf_MVT} yields the distance bound for STA 
\begin{equation}
z_f \ \ge\ \frac{3}{2g_{\max}}\ln\!\left(\frac{a_c(0)}{a_f}\right).
\label{eq:STA_bound_gmax}
\end{equation}
For example, when $g_{\max} =1$, $a_c(0)=1$ and $a_c(z_f)=0.3$, $z_f \ \ge 1.8$. This result makes explicit that, without a physical constraint on the gain/loss amplitude (or equivalently on the management rate),
the formal infimum of $z_f$ is zero; in practice, Eq.~\eqref{eq:STA_bound_gmax} provides a convenient lower-bound benchmark for
compression under gain/loss-limited STA protocols. Finally, we note that the gain/loss-limited bound~\eqref{eq:STA_bound_gmax} is consistent with the adiabatic benchmark distance
\eqref{eq:adiabatic_distance}. In particular, for constant gain $g(z)=g_0$ one has $g_{\max}=g_0$ and the inequality in
\eqref{eq:STA_bound_gmax} is saturated, recovering exactly $z_f^{(\mathrm{adiab})}$. This shows that adiabatic compression corresponds to the ``slowest'' protocol compatible with a fixed gain amplitude, whereas STA achieves shorter distances only by allowing larger peak $|g(z)|$ (and typically introducing gain-loss segments). In this sense, Eq.~\eqref{eq:STA_bound_gmax} quantifies the practical
``cost'' of STA: faster compression requires stronger gain/loss modulation.

\bibliography{sample} 
\end{document}